\documentclass[aps,showpacs,preprintnumbers,amsmath,amssymb]{revtex4}
\usepackage{epsfig,amsmath,amssymb}

\begin{document}


\title{Parrondo's games as a discrete ratchet}
\author{Ra\'{u}l Toral$^{1}$, Pau Amengual$^{1}$}  
\author{and Sergio Mangioni$^2$}
\affiliation{
$^1$ Instituto Mediterr\'aneo de Estudios Avanzados, IMEDEA (CSIC-UIB),\\ ed. Mateu Orfila, Campus UIB, E-07122 Palma de Mallorca, Spain \\
$^2$ Departament de F{\'\i}sica, Facultad de Ciencias Exactas y Naturales, Universidad Nacional de Mar del Plata, De\'an Funes 3350, 7600 Mar del Plata, Argentina}

\begin{abstract}
We write the master equation describing the Parrondo's games as a consistent discretization of the Fokker--Planck equation for an overdamped Brownian particle describing a ratchet. Our expressions, besides giving further insight on the relation between ratchets and Parrondo's games, allow us to precisely relate the games probabilities and the ratchet potential such that periodic potentials correspond to fair games and winning games produce a tilted potential.
\end{abstract}

\maketitle
The Parrondo's paradox\cite{ha99,ha02} shows that the alternation of two losing games can lead to a winning game. This surprising result is nothing but the translation into the framework of very simple gambling games of the ratchet effect\cite{r02}. In particular, the {\sl flashing ratchet}\cite{ab94,pcpa94} can sustain a particle flux by alternating two relaxational potential dynamics, none of which produces any net flux. Despite that this qualitative relation between the Parrondo paradox and the flashing ratchet has been recognized from the very beginning (and, in fact, it constituted the source of inspiration for deriving the paradoxical games), only very recently there has been some interest in deriving exact relations between both\cite{aa03,hkk02}.

In this paper, we rewrite the master equation describing the evolution of the probabilities of the different outcomes of the games in a way that shows clearly its relation with the Fokker--Planck equation for the flashing ratchet. In this way, we are able to give an expression for the dynamical potentials in terms of the probabilities defining the games, as well as an expression for the current. Similarly, given a ratchet potential we are able to construct the games that correspond to that potential.

The Parrondo's paradox considers a player that tosses different coins such that a unit of ``capital" is won (lost) if heads (tails) show up. Although several possibilities have been proposed\cite{pha00,t01,t03,na02,blp02,m00,mb02}
, in this paper we consider the original and easiest version in which the probability of winning, $p_i$, depends on the actual value of the capital, $i$, modulus a given number $L$. A game is then completely specified by giving the set or probabilities $\{p_0,p_1,\dots,p_{L-1}\}$ from which any other value $p_k$ can be derived as $p_k=p_{k \,{\rm mod}\, L}$. A fair game, one in which gains and losses average out, is obtained if $\prod_{i=0}^{L-1}p_i=\prod_{i=0}^{L-1}(1-p_i)$. The paradox shows that the alternation (either random or periodic) of two fair games can yield a winning game. For instance, the alternation of {\sl game A} defined by $p_i\equiv p=1/2,\, \forall i$, and {\sl game B} defined by $L=3$ and $p_0=1/10\,,p_1=p_2=3/4$ produces a winning game although both $A$ and $B$ are fair games. 

A discrete time $\tau$ can be introduced by considering that every coin toss increases $\tau$ by one. If we denote by $P_i(\tau)$ the probability that at time $\tau$ the capital is equal to $i$, we can write the general master equation
\begin{equation}
P_i(\tau+1)=a_{-1}^i P_{i-1}(\tau)+a_0^iP_i(\tau)+a_1^iP_{i+1}(\tau)
\end{equation}
where $a_{-1}^i$ is the probability of winning when the capital is $i-1$,  $a_{1}^i$ is the probability of losing when the capital is $i+1$, and, for completeness, we have introduced $a_0^i$ as the probability that the capital $i$ remains unchanged (a possibility not considered in the original Parrondo games). Note that, in accordance with the rules described before, we have taken that the probabilities $\{a_{-1}^i,a_0^i,a_1^i\}$ do not depend on time. It is clear that they satisfy:
\begin{equation}
a_{-1}^{i+1}+a_0^i+a_1^{i-1}=1
\end{equation}
which ensures the conservation of probability: $\sum_iP_i(\tau+1)=\sum_i P_i(\tau)$.

It is a matter of straightforward algebra to write the master equation in the form of a continuity equation:
\begin{equation}
P_i(\tau+1)-P_i(\tau)=-\left[J_{i+1}(t)-J_i(t)\right]
\end{equation}
where the current $J_i(\tau)$ is given by:
\begin{equation}
\label{current}
J_i(\tau)=\frac{1}{2}\left[F_i P_i(\tau)+F_{i-1}P_{i-1}(\tau)\right]-\left[D_iP_i(\tau)-D_{i-1}P_{i-1}(\tau)\right]
\end{equation}
and
\begin{equation}
F_i = a_{-1}^{i+1}-a_1^{i-1},\hspace{2.0cm}
D_i = \frac{1}{2}(a_{-1}^{i+1}+a_1^{i-1})
\end{equation}
This form is a consistent discretization of the Fokker--Plank equation\cite{hl84} for a probability $P(x,t)$
\begin{equation}
\frac{\partial P(x,t)}{\partial t}=-\frac{\partial J(x,t)}{\partial x}
\end{equation}
with a current
\begin{equation}
J(x,t)=F(x)P(x,t)-\frac{\partial [D(x)P(x,t)]}{\partial x}
\end{equation}
with general drift, $F(x)$, and diffusion, $D(x)$. If $\Delta t$ and $\Delta x$ are, respectively, the time and space discretization steps, such that $x=i\Delta x$ and $t=\tau\Delta t$, it is clear the identification
\begin{equation}
F_i \longleftrightarrow \frac{\Delta t}{\Delta x} F(i\Delta x), \hspace{1.0truecm}
D_i \longleftrightarrow \frac{\Delta t}{(\Delta x)^2} D(i\Delta x)
\end{equation}
The discrete and continuum probabilities are related by $P_i(\tau)\leftrightarrow P(i\Delta x,\tau\Delta t) \Delta x$ and the continuum limit can be taken by considering that $\displaystyle M=\lim_{\Delta t\to 0,\Delta x\to 0}\frac{(\Delta x)^2}{\Delta t}$ is a finite number. In this case $F_i \leftrightarrow M^{-1}\Delta x
F(i\Delta x)$ and $D_i \leftrightarrow M^{-1}D(i\Delta x)$.

From now on, we consider the case $a_0^i=0$. Since $p_i=a_{-1}^{i+1}$ we have
\begin{equation}
D_i\equiv D=1/2\hspace{2.0truecm} F_i=-1+2p_i
\end{equation}
and the current $J_i(\tau)=-(1-p_i)P_i(\tau)+p_{i-1}P_{i-1}(\tau)$ is nothing but the probability flux from $i-1$ to $i$.

The stationary solutions $P_i^{st}$ can be found solving the recurrence relation derived from (\ref{current}) for a constant current $J_i=J$ with the boundary condition $P_i^{st}=P_{i+L}^{st}$: 
\begin{equation}
P_i^{st}=N {\rm e}^{-V_i/D}\left[1- \frac{2 J}{N} \sum_{j=1}^i \frac{{\rm e}^{V_j/D}}{1-F_j}\right]
\end{equation}
with a current
\begin{equation}
\label{corriente}
J=N\frac{{\rm e}^{-V_L/D}-1}{2\sum_{j=1}^L \frac{{\rm e}^{V_j/D}}{1-F_j}}
\end{equation}

$N$ is the normalization constant obtained from $\sum_{i=0}^{L-1} P_i^{st}=1$.
In these expressions we have introduced the potential $V_i$ in terms of the probabilities of the games\footnote{In this, as well as in other similar expressions, the notation is such that $\sum_{j=1}^0=0$. Therefore the potential is arbitrarily rescaled such that $V_0=0$.}
\begin{equation}
\label{potential}
V_i=-D\sum_{j=1}^i\ln\left[\frac{1+F_{j-1}}{1-F_j}\right]=-D\sum_{j=1}^i\ln\left[\frac{p_{j-1}}{1-p_j}\right]
\end{equation}
The case of zero current $J=0$, implies a periodic potential $V_L=V_0=0$. This reproduces again the condition $\prod_{i=0}^{L-1}p_i=\prod_{i=0}^{L-1}(1-p_i)$ for a fair game. In this case, the stationary solution can be written as the exponential of the potential $P_i^{st}=N{\rm e}^{-V_i/D}$. Note that Eq.(\ref{potential}) reduces in the limit $\Delta x\to 0$ to $V(x)=-M^ {-1}\int F(x)dx$ or $F(x)=-M\frac{\partial V(x)}{\partial x}$, which is the usual relation between the drift $F(x)$ and the potential $V(x)$ with a mobility coefficient $M$.

The inverse problem of obtaining the game probabilities in terms of the potential requires solving Eq. (\ref{potential}) with the boundary condition $F_0=F_{L}$\footnote{The singularity appearing for a fair game $V_L=V_0$ in the case of an even number $L$ might be related to the lack of ergodicity explicitely shown in \cite{hkk02} for $L=4$}:
\begin{equation}
\label{fiv}
F_i=(-1)^i{\rm e}^{V_i/D}\left[\frac{\sum_{j=1}^{L}(-1)^j[{\rm e}^{-V_j/D}-{\rm e}^{-V_{j-1}/D}]}{(-1)^L{\rm e}^{(V_0-V_L)/D}-1}+\sum_{j=1}^{i}(-1)^j[{\rm e}^{-V_j/D}-{\rm e}^{-V_{j-1}/D}]\right]
\end{equation}

\begin{figure}[!ht]
\centerline{\makebox{\epsfig{figure=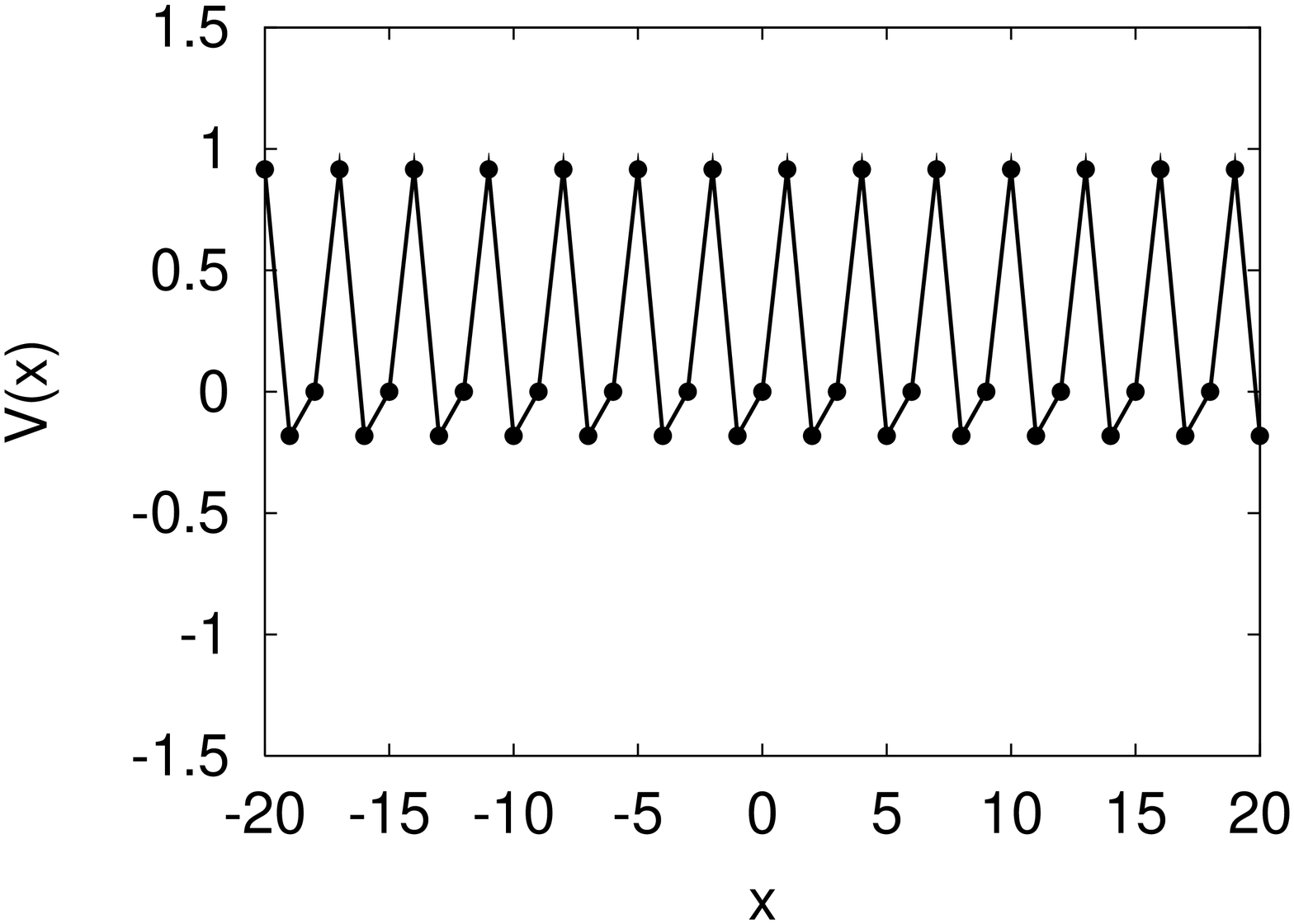,width=5.5cm,height=7cm,angle=-90}\epsfig{figure=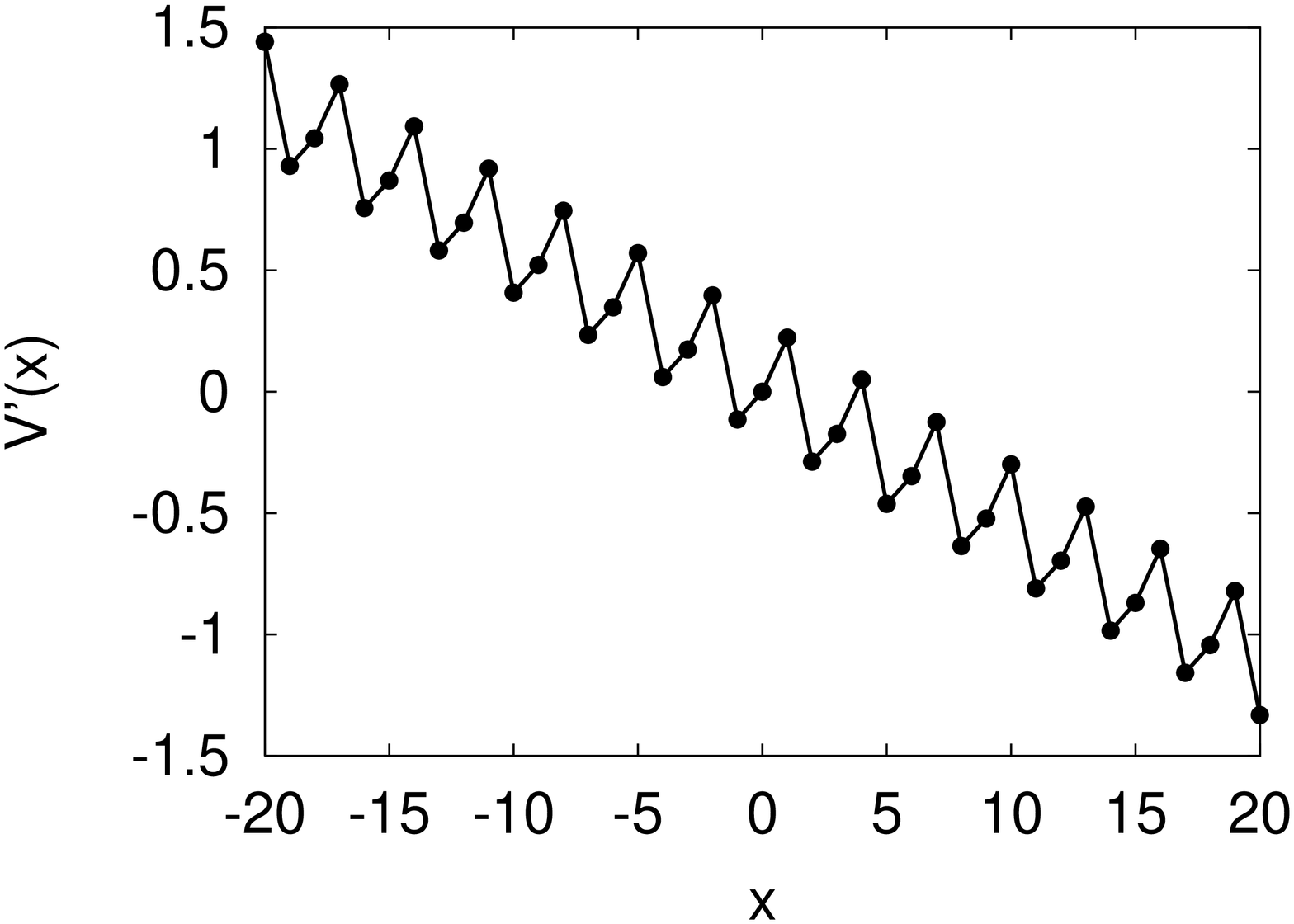,width=5.5cm,height=7cm,angle=-90}}}
\vspace{0.5truecm}\caption{\label{fig1} Left panel: potential $V_i$ obtained from (\ref{potential}) for the fair game $B$ defined by $p_0=1/10,\,p_1=p_2=3/4$. Right panel: potential for game $B'$, with $p_0'=3/10,\,p_1'=p_2'=5/8$ resulting from the random alternation of game $B$ with a game $A$ with constant probabilities $p_i=p=1/2,\,\forall i$.}
\end{figure}

These results allow us to obtain the stochastic potential $V_i$ (and hence the current $J$) for a given set of probabilities $\{p_0,\dots,p_{L-1}\}$, using (\ref{potential}); as well as the inverse: obtain the probabilities of the games given a stochastic potential, using (\ref{fiv}). Note that the game resulting from the alternation, with probability $\gamma$, of a {\sl game A} with $p_i=1/2,\,\forall i$ and a game $B$ defined by the set $\{p_0,\dots,p_{L-1}\}$ has a set of probabilities $\{p'_0,\dots,p'_{L-1}\}$  with $p'_i=(1-\gamma)\frac{1}{2}+\gamma p_i$. For the $F_i$'s variables, this relation yields:
\begin{equation}
F_i'=\gamma F_i,
\end{equation}
and the related potential $V'$ follows from (\ref{potential}). 

We give now two examples of the application of the above formalism. In the first one we compute the stochastic potentials of the fair game $B$ and the winning game $B'$, the random combination with probability $\gamma=1/2$ of game $B$ and a game $A$ with constant probabilities, in the original version of the paradox\cite{ha99}. The resulting potentials are shown in figure \ref{fig1}. Notice how the  potential of the combined game clearly displays the asymmetry under space translation that gives rise to the winning game. 

\begin{figure}[!ht]
\centerline{\makebox{\epsfig{figure=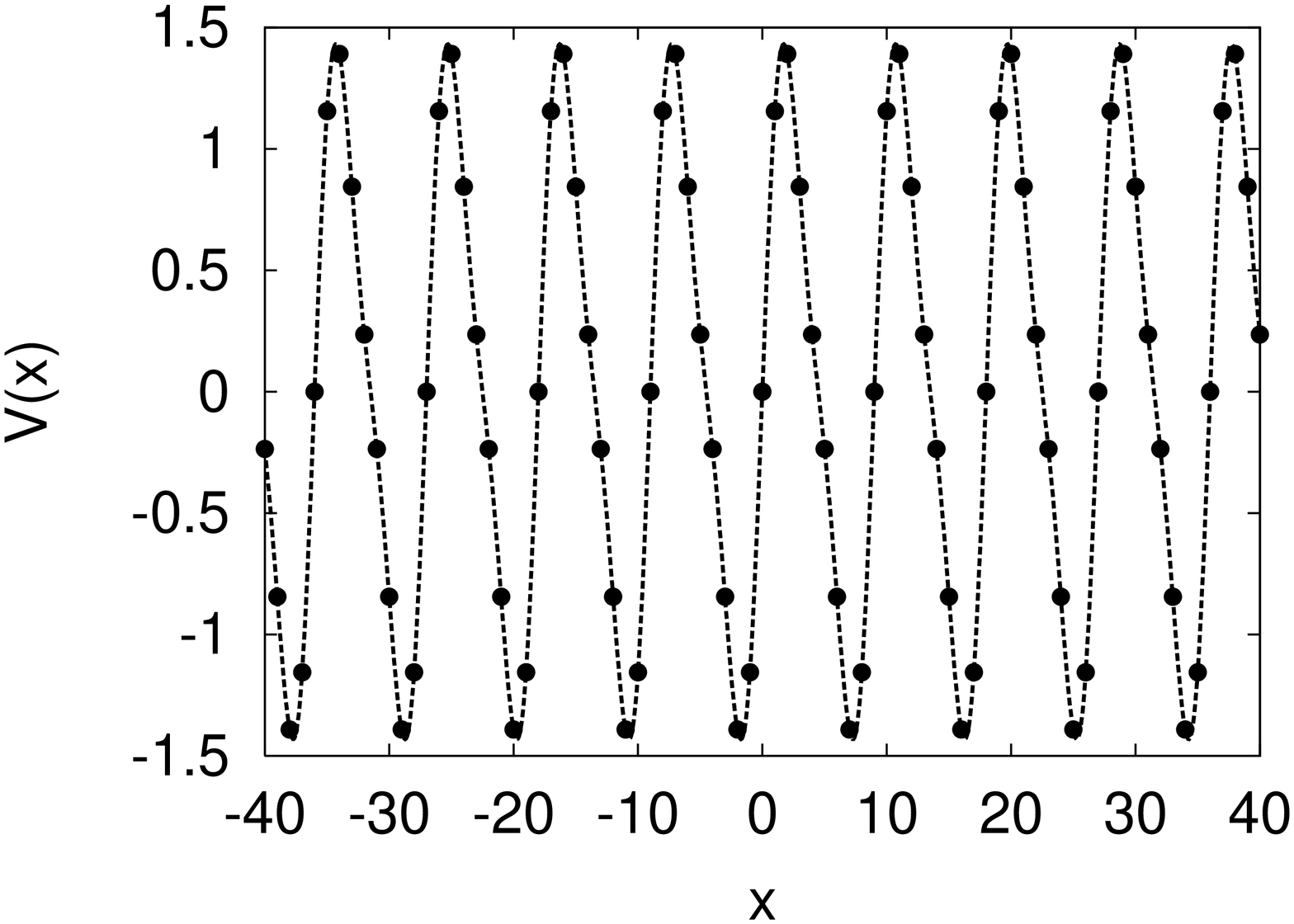,width=5.5cm,height=7cm,angle=-90}\epsfig{figure=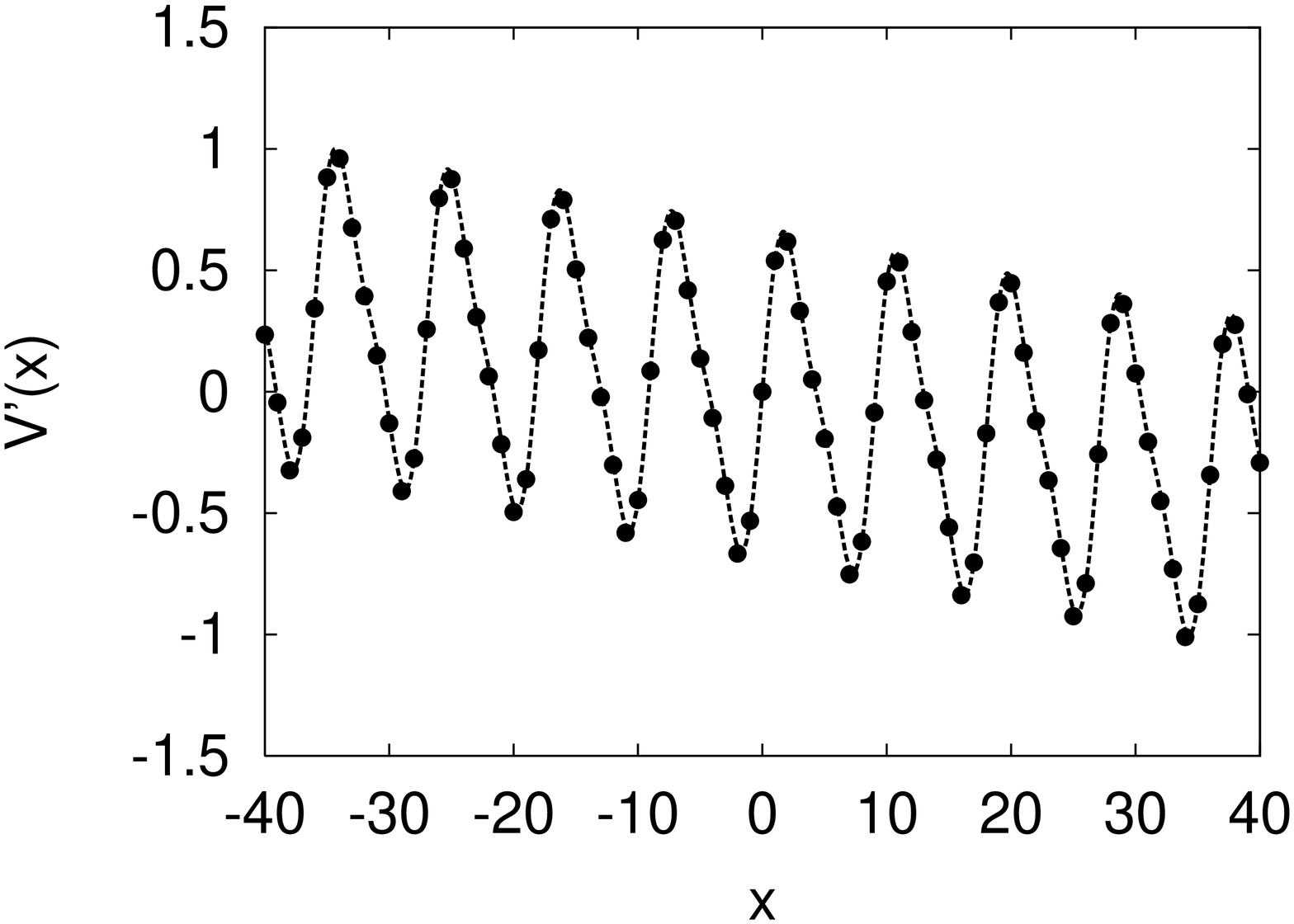,width=5.5cm,height=7cm,angle=-90}}}
\vspace{0.5truecm}\caption{\label{fig2} Left panel: Ratchet potential (\ref{ratchet}) in the case $L=9$, $A=1.3$. The dots are the discrete values $V_i=V(i)$ used in the definition of game $B$. Right panel: discrete values for the potential $V'_i$ for the combined game $B'$ obtained by alternating with probability $\gamma=1/2$ games $A$ and $B$. The line is a fit to the empirical form $V'(x)=-\Gamma x +\alpha V(x)$ with $\Gamma=0.009525$, $\alpha=0.4718$.}
\end{figure}

\begin{figure}[!ht]
\centerline{\makebox{\epsfig{figure=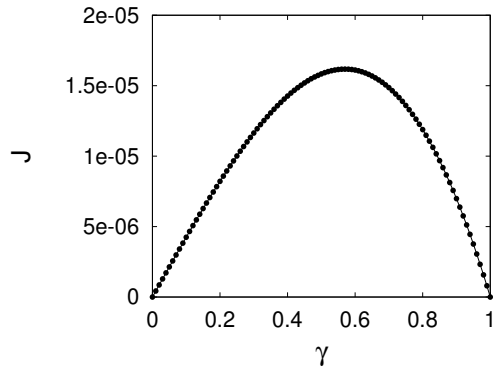,width=5cm,height=7cm,angle=-90}}}
\vspace{0.5truecm}\caption{\label{fig3} Current $J$ resulting from equation (\ref{corriente}) for the game $B'$ as a function of the probability $\gamma$ of alternation of games $A$ and $B$. Game $B$ is defined as the discretization of the ratchet potential (\ref{ratchet}) in the case $A=0.4$, $L=9$. The maximum gain corresponds to $\gamma=0.57$.}
\end{figure}

The second application considers as input the potential 
\begin{equation}
\label{ratchet}
V(x)=A\left[\sin\left(\frac{2\pi x}{L}\right)+\frac{1}{4}\sin\left(\frac{4\pi x}{L}\right)\right]
\end{equation}
which has been widely used as a prototype for ratchets\cite{r02}. Using (\ref{fiv}) we obtain a set of probabilities $\{p_0,\dots,p_{L-1}\}$ by discretizing this potential with $\Delta x=1$, i.e. setting $V_i=V(i)$. Since the potential $V(x)$ is periodic, the resulting game $B$ defined by these probabilities is a fair one and the current $J$ is zero. Game $A$, as always is defined by $p_i=p=1/2,\,\forall i$. We plot in figure \ref{fig2} the potentials for game $B$ and for the game $B'$, the random combination with probability $\gamma=1/2$ of games $A$ and $B$. Note again that the potential $V'_i$ is tilted as corresponding to a winning game $B'$. As shown in figure \ref{fig3}, the current $J$ depends on the probability $\gamma$ for the alternation of games $A$ and $B$. 

In summary, we have written the master equation describing the Parrondo's games as a consistent discretization of the Fokker--Planck equation for an overdamped Brownian particle. In this way we can relate the probabilities of the games $\{p_0,\dots,p_{L-1}\}$ to the dynamical potential $V(x)$. Our approach yields a periodic potential for a fair game and a tilted potential for a winning game. The resulting expressions, in the limit $\Delta x\to 0$ could be used to obtain the effective potential for a flashing ratchet as well as its current.

The work is supported by MCyT (Spain) and FEDER, projects BFM2001-0341-C02-01,
BMF2000-1108. P.A. is supported by a grant from the Govern Balear.

\end{document}